\documentclass[amsmath,amssymb,aps,letterpaper,prl,twocolumn,longbibliography,10pt]{revtex4-1}
\pdfoutput=1
\usepackage{graphicx}
\usepackage{color}
\usepackage{pgfplots}
\pgfplotsset{compat=1.13}

\usepackage{comment}

\usepackage{subfigure}

\usepackage{color}

\makeatletter \let\@keywords\@empty \let\@subject\@empty
\providecommand{\keywords}[1]{\gdef\@keywords{#1}}
\providecommand{\subject}[1]{\gdef\@subject{#1}}
\def\thetitle{\@title}
\def\theauthor{\@author}
\def\thesubject{\@subject}
\def\thedate{\@date}
\def\thekeywords{\@keywords}
\makeatother
\AtBeginDocument{%
\hypersetup{pdftitle={\thetitle}}%
\hypersetup{pdfauthor={\theauthor}}%
\hypersetup{pdfsubject={\thesubject}}%
\hypersetup{pdfkeywords={\thekeywords}}%
}

\usepackage[bookmarks=true,hyperfigures=true]{hyperref}
\usepackage{fixmath}
\usepackage{mathtools}

\providecommand{\href}[2]{#2}

\let\oldbfseries=\bfseries
\let\oldmdseries=\mdseries
\let\oldnormalfont=\normalfont
\renewcommand{\bfseries}{\oldbfseries\boldmath}
\renewcommand{\mdseries}{\oldmdseries\unboldmath}
\renewcommand{\normalfont}{\oldnormalfont\unboldmath}

\makeatletter
\newlength{\apb@width}
\newcommand{\autoparbox}[2][c]{\settowidth{\apb@width}{#2}\parbox[#1]{\apb@width}{#2}}

\makeatother

\DeclareMathOperator{\arctanh}{arctanh}
\newcommand{\e}{\operatorname{e}}

\DeclareMathOperator{\phaneq}{\phantom{{}=}}

\newcommand{\Pb}{\mathbf{P}}
\newcommand{\Qb}{\mathbf{Q}}

\newcommand{\CN}{\mathcal{N}}
\newcommand{\CO}{\mathcal{O}}
\newcommand{\x}{\mathrm{x}}
\newcommand{\y}{\mathrm{y}}

\newcommand{\sep}{\,, \ \ }

\begin{document}

\title{The Hagedorn temperature of \texorpdfstring{AdS$_5$/CFT$_4$}{AdS5/CFT4} at finite coupling \texorpdfstring{\\}{} via the Quantum Spectral Curve}

\author{Troels Harmark}%
 \email{harmark@nbi.ku.dk}
\author{Matthias Wilhelm}%
  \email{matthias.wilhelm@nbi.ku.dk}

\affiliation{%
Niels Bohr Institute, Copenhagen University, 
Blegdamsvej 17, 2100 Copenhagen \O{}, Denmark
}%

\begin{abstract}
Building on the recently established connection between the Hagedorn temperature and integrability \cite{Harmark:2017yrv}, 
we show how the Quantum Spectral Curve formalism can be used to calculate the Hagedorn temperature of AdS$_5$/CFT$_4$ for any value of the 't Hooft coupling.
We solve this finite system of finite-difference equations perturbatively at weak coupling and numerically at finite coupling. We confirm previous results at weak coupling and obtain the previously unknown three-loop Hagedorn temperature.
Our finite-coupling results interpolate between weak and strong coupling and allow us to extract the first perturbative order at strong coupling. Our results indicate that the Hagedorn temperature for large 't Hooft coupling approaches that of type IIB string theory in ten-dimensional Minkowski space. 
\end{abstract}

\maketitle

\section{Introduction}

The $\mbox{AdS}_5$/$\mbox{CFT}_4$ correspondence \cite{Maldacena:1997re} provides an exact duality between two seemingly very different theories. On the one side, one has a four-dimensional gauge theory in the form of $\CN=4$ super-Yang-Mills (SYM) theory on $\mathbb{R}\times S^3$ with 't Hooft coupling $\lambda$. On the other side, one has type IIB string theory on the ten-dimensional target-space AdS$_5 \times S^5$. This makes the $\mbox{AdS}_5$/$\mbox{CFT}_4$ correspondence an important theoretical laboratory for understanding various interesting problems in physics. One such problem is the nature of the Hagedorn temperature in string theory. Tree-level string theory has an exponentially growing density of states at large energies, which leads to a singularity in the thermodynamic partition function defining the Hagedorn temperature.
Since the $\mbox{AdS}_5$/$\mbox{CFT}_4$ correspondence provides a non-perturbative definition of string theory, it should enable one to study the Hagedorn temperature and all its related phenomena.

In the $\mbox{AdS}_5$/$\mbox{CFT}_4$ correspondence, the Hagedorn singularity is connected to the Hawking-Page transition that occurs at a lower temperature where the black hole phase becomes thermodynamically favorable over a gas of closed strings. On the gauge-theory side, this corresponds to the confinement-deconfinement transition, where the confined phase occurs due to the confinement of the color degrees of freedom on a three-sphere \cite{Atick:1988si, Witten:1998zw,Sundborg:1999ue,Aharony:2003sx}. 
However, in the limit of zero string coupling, or the strict limit of infinite colors on the gauge-theory side, this transition requires arbitrarily high energy to realize, and one is left with the Hagedorn temperature as a maximal possible temperature on both sides of the correspondence. In this somewhat simpler setting, a starting point for further exploration of finite-temperature physics is to establish a quantitative interpolation of the Hagedorn temperature between the gauge-theory and string-theory sides.

However, it is only for certain precious cases that one has exact methods available to make a quantitative interpolation from weak to strong 't Hooft coupling. One such method is integrability, see Refs.\ \cite{Beisert:2010jr,Bombardelli:2016rwb} for reviews. Recently, we proposed a framework for calculating the Hagedorn temperature of $\mbox{AdS}_5$/$\mbox{CFT}_4$ using integrability \cite{Harmark:2017yrv}. In this Letter, we take this a step further by exploiting this connection to compute the Hagedorn temperature at finite 't Hooft coupling. This enables us to interpolate all the way from zero 't Hooft coupling to large 't Hooft coupling where we find the Hagedorn temperature of type IIB string theory in flat space, in both cases matching a previous computation of Sundborg \cite{Sundborg:1999ue,Sundborg:1984uk}.

The proposal \cite{Harmark:2017yrv} for calculating the Hagedorn temperature of $\mbox{AdS}_5$/$\mbox{CFT}_4$ via integrability is as follows. 
Define $F(T)$ to be the free energy per unit classical scaling dimension in the limit of large classical scaling dimension of the spin chain associated with planar $\CN=4$ SYM theory. Then the Hagedorn temperature $T_{\rm H}$ in units of the $S^3$ radius is determined by
\begin{equation}
\label{eq: Hagedorn eq}
F(T_{\rm H}) = -1 \,,
\end{equation}
for zero chemical potentials. This can be seen from the fact that Eq.\ \eqref{eq: Hagedorn eq} determines the temperature beyond which the planar partition function of $\mathcal{N}=4$ SYM theory is singular.
The free energy is computed from so-called Thermodynamic Bethe Ansatz (TBA) equations \cite{Harmark:2017yrv}, which are an infinite system of integral equations. In Ref.\ \cite{Harmark:2017yrv}, we have solved these equations perturbatively at weak coupling, reproducing the know tree-level result \cite{Sundborg:1999ue} and one-loop correction \cite{Spradlin:2004pp} as well as finding the previously unknown two-loop correction.
In principle, one can employ the TBA equations to compute higher order corrections to $T_{\rm H}$ and to find $T_{\rm H}$ numerically at finite coupling as well.
In practice, however, the nature of these equations massively complicates perturbative calculations and severely limits the numeric accuracy one can achieve at finite coupling.

The TBA equations for the spectral problem of $\mathcal{N}=4$ SYM theory were recast into the form of the Quantum Spectral Curve (QSC) \cite{Gromov:2013pga,Gromov:2014bva,Gromov:2014caa}, see Refs.\ \cite{Gromov:2017blm,Kazakov:2018ugh} for reviews.
It consists of a finite system of finite-difference equations, which allows for a very efficient evaluation both perturbatively at weak coupling \cite{Marboe:2014gma,Marboe:2014sya,Marboe:2017dmb} and numerically at finite coupling \cite{Gromov:2015wca,Hegedus:2016eop}.
The QSC was since used for 
the pomeron \cite{Alfimov:2014bwa,Gromov:2015vua},
cusped Wilson lines and the quark-antiquark potential \cite{Gromov:2015dfa,Gromov:2016rrp,Cavaglia:2018lxi}
as well as 
integrable deformations of $\mathcal{N}=4$ SYM theory \cite{Kazakov:2015efa,Klabbers:2017vtw,Gromov:2017cja}.

In this Letter, we recast our TBA equations for the Hagedorn temperature $T_{\rm H}$
of $\mbox{AdS}_5$/$\mbox{CFT}_4$ into the form of the QSC.
Moreover, we solve these equations perturbatively at weak coupling and numerically at finite coupling.
%

\section{QSC equations for the Hagedorn temperature}

\paragraph{QSC equations}

The TBA equations are an infinite system of integral equations given in terms of Y-functions.
They can be recast into the form of the so-called Y-system and T-system, which are infinite systems of finite-difference equations, and subsequently into the form of the so-called Q-system, which is a finite system of finite-difference equations also known as the Quantum Spectral Curve (QSC). Since we are setting all chemical potentials to zero, we are in a situation with so-called left-right symmetry, which is a symmetry between the two $\mathfrak{su}(2|2)$ subalgebras of the superconformal symmetry algebra $\mathfrak{psu}(2,2|4)$.
The QSC is then formulated in terms of the functions $\Pb_a(u)$, $\Qb_i(u)$ and $Q_{a|i}(u)$, where $a,i=1,2,3,4$ and $u$ is the spectral parameter. They satisfy the finite-difference equations
\begin{eqnarray}
\label{eq: QSC equation1}
&Q_{a|i}^+ - Q_{a|i}^- = \Pb_a \Qb_i \,, &
\\
\label{eq: QSC equation2}
&\Pb_a = - \Qb^i Q^+_{a|i} \,,&
\end{eqnarray}
where $f^\pm (u) = f(u\pm \tfrac{i}{2})$. The functions $Q_{a|i}$ are orthonormal in the sense that 
\begin{equation}
\label{eq: QSC equation3}
Q_{a|i}Q^{b|i}=-\delta_a^b\,,\qquad Q_{a|i}Q^{a|j}=-\delta_i^j \,.
\end{equation} 
Here the functions with upper indices are defined as
\begin{equation}
\label{eq: QSC upper indices}
\Pb^a = \chi^{ab} \Pb_b \,, \ \  \Qb^i = \chi^{ij} \Qb_j \,, \ \ Q^{a|i} = \chi^{ab} \chi^{ij} Q_{b|j} \,,
\end{equation}
where the non-zero entries of $\chi$ are $\chi^{14}=\chi^{32} = - 1$, $\chi^{23}=\chi^{41}=1$.
The Eqs.~\eqref{eq: QSC equation1}--\eqref{eq: QSC equation3} reflect the $\mathfrak{psu}(2,2|4)$ symmetry of $\mathcal{N}=4$ SYM theory, where $\Pb_a(u)$ ($\Qb_i(u)$) is associated to the conformal symmetry $\mathfrak{su}(2,2)$ (R-symmetry $\mathfrak{su}(4)$).
They are universal in the sense that they do not depend on the specific physical observable that one is computing
but are common to all cases so far investigated.
In order to specify a particular physical observable, these universal equations have to be supplemented by the asymptotic behavior of the functions at large spectral parameter $u$, by the location of the branch cuts and by the discontinuities across these branch cuts.

\paragraph{Asymptotic behavior}

We can infer the asymptotic behavior of $\Pb_a(u)$ and $\Qb_i(u)$ at large spectral parameter $u$ from the asymptotic behavior of the Y-functions found in Ref.\ \cite{Harmark:2017yrv}. At large $u$, the Y-functions asymptote to constants determined from a one-parameter family of constant T-systems with parameter $z$, see Eqs.~(26)--(27) in Ref.\ \cite{Harmark:2017yrv}. Using the TBA equations, we show in Ref.\ \cite{Harmark:2021qma} that $F(T) = - 4T \arctanh z$. Since the Hagedorn temperature $T_{\rm H}$ satisfies Eq.~\eqref{eq: Hagedorn eq}, this fixes
$z = \tanh\frac{1}{4T_{\rm H}}$.
The asymptotic Y-functions are reproduced via 
\footnote{Interestingly, the asymptotics \eqref{eq: asymptotic P}--\eqref{eq: asymptotic Q} are similar to the ones of the spectral problem of twisted $\mathcal{N}=4$ SYM theory \cite{Kazakov:2015efa}. Concretely, excluding the prefactors they are formally the same if we set the twists $1/\x_1=1/\x_2=\x_3=\x_4=-\e^{-\frac{1}{2T_{\text{H}}}}$, $\y_1=\y_2=\y_3=\y_4=1$ and insert the Cartan charges $\Delta=S_1=S_2=J_1=J_2=J_3=0$.
This is related to how the constant T-system \cite{Harmark:2017yrv} is obtained from the general $\mathfrak{psu}(2,2|4)$ character solution \cite{Gromov:2010vb}.}
\begin{equation}
\label{eq: asymptotic P}
 \begin{aligned}
  &\Pb_{1}(u) = A_1 \big(-\e^{-\frac{1}{2T_{\text{H}}}}\big)^{-i u}\left(1+\mathcal{O}(u^{-1})\right),\\
  &\Pb_{2}(u) = A_2 \big(-\e^{-\frac{1}{2T_{\text{H}}}}\big)^{-i u}\left(u+\mathcal{O}(u^{0})\right),\\
  &\Pb_{3}(u) = A_3 \big(-\e^{-\frac{1}{2T_{\text{H}}}}\big)^{+i u}\left(1+\mathcal{O}(u^{-1})\right),\\
  &\Pb_{4}(u) = A_4 \big(-\e^{-\frac{1}{2T_{\text{H}}}}\big)^{+i u}\left(u+\mathcal{O}(u^{0})\right)
 \end{aligned}
\end{equation}
and
 \begin{equation}
\label{eq: asymptotic Q}
   \begin{aligned}
  &\Qb_{1}(u)=B_1 \left(1+\mathcal{O}(u^{-1})\right),&
 &  \Qb_{2}(u) =B_2 \left(u+\mathcal{O}(u^{0})\right),\\
 &\Qb_{3}(u)= B_3 \left(u^2+\mathcal{O}(u^{1})\right),&
 & \Qb_{4}(u)=B_4 \left(u^3+\mathcal{O}(u^{2})\right), 
 \end{aligned}
 \end{equation}
with 
$
A_1A_4=A_2A_3=\frac{i}{\tanh^2 \frac{1}{4T_{\text{H}}}}
$
and 
$
3B_1B_4=B_2B_3=-8 i\cosh^4 \frac{1}{4T_{\text{H}}}
$.
The asymptotic behavior of $Q_{a|i}(u)$ then follows from Eq.\ \eqref{eq: QSC equation1}.

\paragraph{Branch cut structure and ansatz}

We consider the so-called direct theory rather than the mirror theory; hence, we use the Zhukowski variable
\begin{equation}
x(u) = \frac{u}{2g} \left( 1 + \sqrt{1-\frac{4g^2}{u^2}} \right) \,,
\end{equation}
which has a `short' branch cut  at the interval $(-2g,2g)$, where $g^2=\frac{\lambda}{16\pi^2}$ is the effective planar loop coupling.
We work in a Riemann sheet in which the 
four functions $\Qb_i(u)$ have one short cut at the interval $(-2g,2g)$, while $\Pb_a(u)$ and its analytic continuation $\tilde\Pb_a(u)$ have an infinite set of short cuts at $(-2g,2g) - in$ and $(-2g,2g) + in$ with $n \in \mathbb{N}_{\geq 0}$, respectively.
Since $\Qb_i(u)$ has only a single short cut, we can make the ansatz
\begin{equation}
\label{eq: ansatz}
\begin{aligned}
\Qb_{i} (u) = B_i (gx(u))^{i-1} \left( 1+\sum_{n=1}^\infty \frac{c_{i,n}(g)}{(gx(u))^{2n}} \right)\,.
\end{aligned}
\end{equation}
For convenience, we choose the gauge $c_{3,1}=0$ and  $B_1=B_2 = 1$.

The previous applications of the QSC formalism have all been in the mirror theory rather than in the direct theory that we consider here. In the mirror theory, it is the four functions $\Pb_a(u)$ for which one can choose a Riemann sheet where they have only one short cut at the real axis. This means one makes an ansatz for the functions $\Pb_a(u)$ instead of the functions $\Qb_i(u)$ as we do in our case.

\paragraph{Gluing conditions}

To close the system of QSC equations, one needs to impose so-called gluing conditions \cite{Gromov:2017blm}. They relate the analytic continuation $\tilde{\Pb}_a(u)$ through the short cut at the real axis to a linear combination of the complex conjugates of the $\Pb_b(u)$ functions.
In our case, using the gauge choice $A_1=iA_2=-A_3=-iA_4=(\tanh\frac{1}{4T_{\text{H}}})^{-1}$
for the asymptotics \eqref{eq: asymptotic P}, the gluing conditions are
\begin{equation}
\label{eq: gluing conditions}
 \tilde\Pb_a(u)=(-1)^{1+a}\overline{\Pb_a(u)}\,.
\end{equation}

Together with the QSC equations \eqref{eq: QSC equation1}--\eqref{eq: QSC equation3}, the ansatz \eqref{eq: ansatz} for the functions $\Qb_i(u)$ and the large-$u$ asymptotics \eqref{eq: asymptotic P}--\eqref{eq: asymptotic Q},
the gluing conditions \eqref{eq: gluing conditions} determine the Hagedorn temperature $T_{\rm H}$ for any given value of $g$ -- which is one of the main results of this Letter. Another main result is that we 
 will explicitly solve these equations perturbatively at weak coupling and numerically at finite coupling, as explained below.

\section{Perturbative solution}

To solve the QSC equations perturbatively at weak coupling, we start with the tree-level solution for $g=0$. 
It can be obtained from Eqs.\ \eqref{eq: asymptotic P}--\eqref{eq: asymptotic Q} by setting $T_{\text{H}}$ to the tree-level Hagedorn temperature $T_{\text{H}}^{(0)}=1/(2\log(2+\sqrt{3}))$ \cite{Sundborg:1999ue},
which determines the leading coefficients.
The only non-vanishing subleading coefficient at tree level is $c_{4,1}(0)=-1$. 
Using the Eqs.\ \eqref{eq: QSC equation1}--\eqref{eq: QSC equation3}, one finds the corresponding functions $Q_{a|i}(u)$ at tree level, which we denote below as $Q_{a|i}^{(0)}(u)$. 

Knowing the tree-level solution, we can now solve the QSC equations \eqref{eq: QSC equation1}--\eqref{eq: QSC equation3} perturbatively following a slightly modified version of the approach in Ref.\ \cite{Gromov:2015vua}. 
Write the solution of Eq.~\eqref{eq: QSC equation1} as
\begin{equation}
 Q_{a|i}=Q_{a|i}^{(0)}+(b_a{}^c) {}^+Q_{c|i}^{(0)}\,.
\end{equation}
Then $b_a{}^c(u)$ satisfies the first order finite-difference equation
\begin{equation}
\label{eq: b equation}
 (b_a{}^c) {}^{++}-b_a{}^c=dS_{a|i}(Q^{(0)c|i}) {}^-+(b_a{}^b) {}^{++} dS_{b|i}(Q^{(0)c|i}) {}^-\,,
\end{equation}
where $dS_{a|i}$ is defined as
\begin{equation}
 dS_{a|i}\equiv Q_{a|i}^{(0)} {}^+ -   Q_{a|i}^{(0)} {}^-    +  \Qb_i\Qb^j  Q_{a|j}^{(0)} {}^+\,.
\end{equation}
Here, we use the ansatz \eqref{eq: ansatz} for $\Qb_i(u)$, in which the sum truncates for any given loop order $\ell$. Assume one has already determined the coefficients of $\Qb_i(u)$ in Eq.\ \eqref{eq: ansatz} at $(\ell-1)$-loop order.
Solving Eq.~\eqref{eq: b equation}, we find $b_a{}^c(u)$ and hence $Q_{a|i}(u)$ at $\ell$-loop order in terms of the as yet undetermined parameters in the ansatz \eqref{eq: ansatz} as well as additional undetermined constant parameters from the homogeneous solution to Eq.~\eqref{eq: b equation}. Because of the phases in Eq.\ \eqref{eq: asymptotic P}, we encounter finite-difference equations of the type
\begin{equation}
z_i f^{++}(u) - f(u) = h(u) \sep z_i \in \{ 1, (2+\sqrt{3})^{\pm 2} \}\,.
\end{equation}
The solution can be written in terms of the generalized $\eta$ functions \cite{Gromov:2015dfa,Kazakov:2015efa}
\begin{equation}
\label{eq: eta definition}
 \eta^{z_1,\dots,z_k}_{s_1,\dots,s_k}(u)\equiv \sum_{n_1>n_2>\dots >n_k\geq 0}\frac{z_1^{n_1}\dots z_k^{n_k}}{(u+i n_1)^{s_1}\dots(u+i n_k)^{s_k}}
 \,.
\end{equation}
Since in some cases $z_i = (2+\sqrt{3})^2$, we have to use analytic continuation as a regularization. Note that this is a worse divergence than in the twisted QSC \cite{Kazakov:2015efa} since in that case $z_i$ is on the unit circle.
Evaluated at $u=i$, the generalized $\eta$ functions are proportional to multiple polylogarithms
\begin{equation}
 \text{Li}_{s_1,\dots,s_k}(z_1,\dots,z_k)\equiv \sum_{n_1>n_2>\dots >n_k> 0}\frac{z_1^{n_1}\dots z_k^{n_k}}{ n_1^{s_1}\dots n_k^{s_k}}
 \,,
\end{equation}
in terms of which the result for the Hagedorn temperature is naturally expressed.
Note that the multiple polylogarithms have branch cuts on the real axis; for instance, classical polylogarithms have branch cuts between $1$ and $\infty$.
The ambiguity in evaluating these polylogarithms on the branch cut can be resolved by an $i \epsilon$ prescription.

The next step is to impose conditions to determine the unfixed parameters.
To begin with, we impose Eq.~\eqref{eq: QSC equation3} which fixes half of the coefficients from the homogeneous solution in $Q_{a|i}(u)$. 
By Eq.~\eqref{eq: QSC equation2}, we now find $\Pb_a(u)$.
The gluing conditions \eqref{eq: gluing conditions} enter by imposing that $\Pb_a(u)+\tilde\Pb_a(u)$ and $(\Pb_a(u)-\tilde\Pb_a(u))/\sqrt{u^2-4g^2}$ are regular at $u=0$. 
Finally, we have to impose the asymptotic behavior \eqref{eq: asymptotic P}, which we implement by requiring that 
\begin{equation}
\label{eq: asymptotic P functions}
 \begin{aligned}
\frac{\Pb_2(u)}{\Pb_1(u)} &= -iu + \CO (u^0) \,.
\end{aligned}
\end{equation}
Up to a gauge choice, this fixes all parameters of the ansatz \eqref{eq: ansatz} at $\ell$-loop order, including the Hagedorn temperature $T_{\rm H}$. 
\footnote{In contrast, the $\ell$-loop anomalous dimension in the spectral problem is only fixed at the $(\ell+1)$th order.}

We find, up to three-loop order, 
\begin{widetext}
\begin{equation}
\label{eq: perturbative result}
 \begin{aligned}
  T_{\text{H}}(g)&=\frac{1}{2\log(2+\sqrt{3})}
 +g^2\frac{1}{\log(2+\sqrt{3})} 
 + g^4\Biggl(48 -\frac{86}{ \sqrt{3}} - \frac{48 \text{Li}_1\Big(\tfrac{1}{(2+\sqrt{3})^2}\Big)}{\log(2+\sqrt{3})}\Biggr)
 +g^6\Biggl(
 624 \text{Li}_2\Big(\tfrac{1}{(2+\sqrt{3})^2}\Big)
 \\
 &\phaneq+\frac{432 \text{Li}_1\Big(\tfrac{1}{(2+\sqrt{3})^2}\Big)^2}{\log (2+\sqrt{3})}+\frac{312 \text{Li}_3\Big(\tfrac{1}{(2+\sqrt{3})^2}\Big)}{\log (2+\sqrt{3})}
+\left(384 \sqrt{3}-864+416 \log (2+\sqrt{3})\right) \text{Li}_1\Big(\tfrac{1}{(2+\sqrt{3})^2}\Big)
 \\&\phaneq-\frac{20}{\sqrt{3}}+\left(\frac{1900}{3}-384 \sqrt{3}\right) \log (2+\sqrt{3})
 \Biggr)
 +\mathcal{O}(g^{8})
 \\
 &\approx(0.3796628588\dots)+(0.7593257175\dots)g^2+(-4.367638556\dots)g^4+(37.22529358\dots)g^6+\mathcal{O}(g^{8})
  \,.
\end{aligned}
\end{equation}
\end{widetext}
This agrees with the previously known results at tree level \cite{Sundborg:1999ue}, one-loop order \cite{Spradlin:2004pp} and two-loop order \cite{Harmark:2017yrv}.
In an upcoming publication \cite{Harmark:2021qma}, we will also present the result at four-loop order and beyond.

\section{Numerical solution}

The QSC can also be solved numerically at finite values of $g$.
Concretely, it can be reduced to a minimization problem that can be solved iteratively.
We use a modified version of the approach in Ref.\ \cite{Gromov:2015wca}, see also Refs.\ \cite{Gromov:2017blm,Alfimov:2018cms}.

Each iteration starts at some values for the Hagedorn temperature $T_{\rm H}$ and the coefficients $c_{1,n}$, $c_{2,n}$ for $n=1,\dots,K$, $c_{3,n}$ for $n=2,\dots,K$ and $c_{4,n}$ for $n=3,\dots,K$ in the ansatz \eqref{eq: ansatz}, which is truncated at some finite $K$.
We make an ansatz for $Q_{a|i}(u)$ at large $u$:
\begin{equation}
Q_{a|i}(u) =  \Big(-\e^{-\tfrac{1}{2T_{\rm H}}}\Big)^{-s_a i u} u^{p_{a|i}}  \sum_{n=0}^{N} \frac{B_{a|i,n}}{u^n}\,,
\end{equation}
which is truncated at some finite order depending on $N$.
Here, we have used
$s_a=1$ and $p_{a|i}=a+i-2$ for $a=1,2$ as well as $s_a=-1$ and $p_{a|i}=a+i-4$ for $a=3,4$. We solve for the remaining coefficients $B_{a|i,n}$ by imposing Eqs.\ \eqref{eq: QSC equation3} and 
\eqref{eq: QSC equation1} where $\Pb_a(u)$ is eliminated using Eq.\ \eqref{eq: QSC equation2}.
In particular, this also fixes $c_{4,1}$ and $c_{4,2}$.

Starting at some finite but large imaginary value of $u$, we can shift $Q_{a|i}$ towards the real axis in steps of $i$ using
\begin{equation}
 Q_{a|i}^- = Q_{a|i}^++ \Qb_i\Qb^j Q^+_{a|j}  \,, 
\end{equation}
which follows from Eqs.\ \eqref{eq: QSC equation1}--\eqref{eq: QSC equation2}.
We can now reconstruct $\Pb_a$ and its analytic continuation
\begin{equation}
 \tilde\Pb_a = - \tilde\Qb^i Q^+_{a|i} 
\end{equation}
on the real axis, where $\tilde\Qb_i$ is obtained from the ansatz \eqref{eq: ansatz} via $\tilde{x}=1/x$. 
Note that this is however only possible for $a=1,2$, as in this case $Q^+_{a|i}$ is exponentially small for large imaginary $u$, while it is exponentially large for $a=3,4$.

Now we can define a function $F$ that vanishes for an exact solution of the gluing conditions \eqref{eq: gluing conditions}:
\begin{equation}
 F(T_{\text{H}},\{c_{i,n}\})=\sum_{a=1}^{2}\sum_{j=1}^P \left|\frac{\overline{\Pb_a(p_i)}}{\tilde\Pb_a(p_i)}+(-1)^{a}\right|^2\,,
\end{equation}
where $p_i$ are $P$ points in the interval $(-2g,2g)$.
We can find an approximate solution for $T_{\text{H}}$ and the coefficients $c_{i,n}$ by minimizing $F$ iteratively, using for instance Newton's method or the Levenberg-Marquardt algorithm.

\begin{figure}
\centering
 \begin{tikzpicture}
  \begin{axis}[
  /pgf/number format/set thousands separator = {},
    xlabel = $g^2$,
    ylabel = $T_{\text{H}}$,
    xmin=0,
    xmax=0.1,
    legend pos = north west,
    ]
    \addplot [only marks, black,mark size=1] table[col sep=comma,x index=0,y index=1,header=false] {datagsquared.csv};
     \addplot [blue,domain=0:0.1,dash pattern=on 5pt off 5pt] {0.3796628587501034616094920614718360257603178667293620488696476738937120349688174722116960880927288187};
     \addplot [red,domain=0:0.1,dash pattern=on 3pt off 3pt on 5pt off 3pt] {0.3796628587501034616094920614718360257603178667293620488696476738937120349688174722116960880927288187+2*x*0.3796628587501034616094920614718360257603178667293620488696476738937120349688174722116960880927288187};
     \addplot [green,domain=0:0.1,dash pattern=on 3pt off 2pt on 1pt off 2pt] {0.3796628587501034616094920614718360257603178667293620488696476738937120349688174722116960880927288187
     +2*x*0.3796628587501034616094920614718360257603178667293620488696476738937120349688174722116960880927288187
     -4.367638555796405907668892892349321686325629017277959319603305534781703926169081640936457034108387146*x^2};
     \addplot [orange,domain=0:0.1,dash pattern=on 3pt off 3pt] {0.3796628587501034616094920614718360257603178667293620488696476738937120349688174722116960880927288187
     +2*x*0.3796628587501034616094920614718360257603178667293620488696476738937120349688174722116960880927288187
     -4.367638555796405907668892892349321686325629017277959319603305534781703926169081640936457034108387146*x^2
     +37.22529357787421243436491861519803356882481034413398637710896967279804321042713467179649224766014504*x^3};
     \addlegendentry{numeric}
     \addlegendentry{0-loop}
     \addlegendentry{1-loop}
     \addlegendentry{2-loop}
     \addlegendentry{3-loop}
  \end{axis}
\end{tikzpicture}
\caption{Numeric results and weak coupling approximation at various loop orders for the Hagedorn temperature as a function of $g^2$.}
\label{fig: weak coupling numerics}
\end{figure}
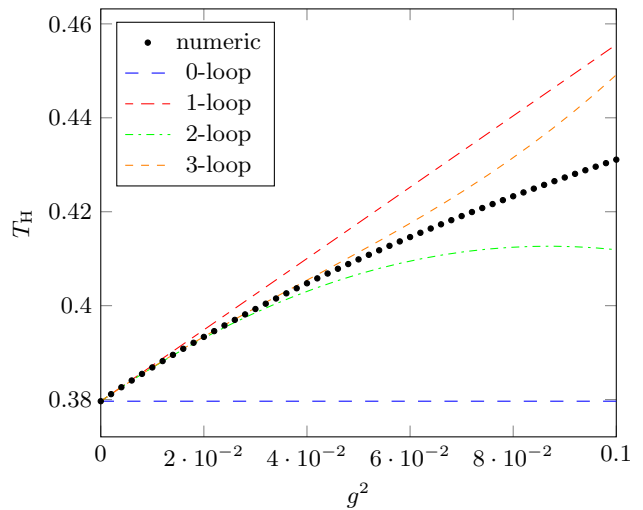

We have plotted our numeric results for the Hagedorn temperature $T_{\text{H}}$ as a function of $g^2$ for $0\leq g^2\leq0.1$ in Fig.\ \ref{fig: weak coupling numerics}.
In addition, Fig.\ \ref{fig: weak coupling numerics} contains the perturbative approximation to the non-perturbative results up to three-loop order. As expected, the perturbative series converges towards the exact results for sufficiently small values of $g^2$.

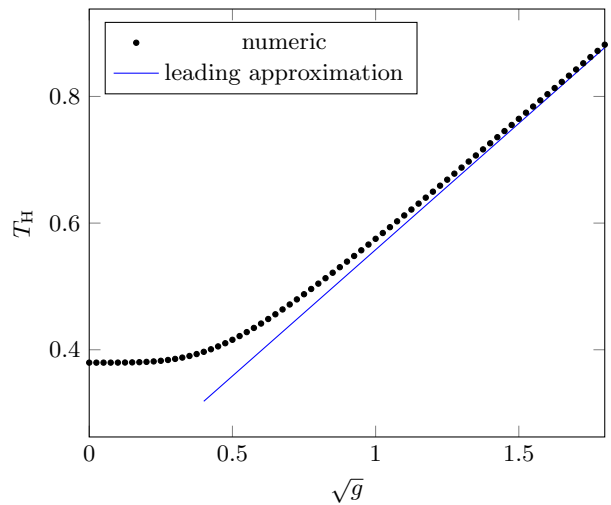
\begin{figure}
\centering
 \begin{tikzpicture}
  \begin{axis}[
  /pgf/number format/set thousands separator = {},
    xlabel = $\sqrt{g}$,
    ylabel = $T_{\text{H}}$,
    xmin=0,
    xmax=1.8,
    legend pos = north west,
    ]
    \addplot [only marks, black,mark size=1] table[col sep=comma,x index=0,y index=1,header=false] {datagsqroot.csv};
    \addplot [blue,domain=0.4:1.8]
    {0.3989422804014326779399460599343818684758586311649346576659258296706579258993018385012523339073069364*x
    +0.159032};
     \addlegendentry{numeric}
     \addlegendentry{leading approximation}
  \end{axis}
\end{tikzpicture}
\caption{Numeric results and leading strong coupling approximation for the Hagedorn temperature as a function of $\sqrt{g}$.}
\label{fig: strong coupling numerics}
\end{figure}
Fig.\ \ref{fig: strong coupling numerics} shows our numeric data for $T_{\text{H}}$ as a function of $\sqrt{g}$ for $0\leq \sqrt{g} \leq1.8$. In particular, we see that $T_{\text{H}}$ tends towards a linear function in $\sqrt g$ at strong coupling.
Using a sixth-order fit in $1/\sqrt{g}$, we find the following approximate result for the leading coefficient:
\begin{equation}
\label{eq: leading approximation}
 T_{\text{H}}(g)=(0.399\dots)\sqrt{g}+\mathcal{O}(g^0)\,,
\end{equation}
where the uncertainty is in the last digit \footnote{%
In a certain exactly solvable pp-wave limit of string theory, the leading contribution to the Hagedorn temperature is of order $\lambda^{1/3}$ with corrections in $1/\lambda^{1/3}$ \cite{Harmark:2006ta}.
In the present case, the leading contribution is of order $\lambda^{1/4}$ and we thus expect the corrections to be in $1/\lambda^{1/4}$.
}.
At the given accuracy, this agrees with the expectation 
that the Hagedorn temperature measured in units of the $S^3$ radius $R$ approaches the behavior
\begin{equation}
\label{tflat1}
T_{\rm H}(g) \simeq \sqrt{\frac{g}{2\pi}}\approx (0.3989422804\dots)\sqrt{g}
\end{equation}
for large 't Hooft coupling. This corresponds to the Hagedorn temperature 
of tree-level type IIB string theory on ten-dimensional Minkowski space \cite{Sundborg:1984uk}. One can see this explicitly by reinstating $1/R$ using the AdS$_5$/CFT$_4$ dictionary. In terms of this, the $S^3$ radius $R$  corresponds on the string-theory side to the radius $R = \lambda^{1/4} l_s$ of AdS$_5$ and $S^5$, where  
$l_s$ is the string length. Hence, Eq.~\eqref{tflat1} becomes
\begin{equation}
\label{tflat2}
T_{\rm H}(g) \simeq \frac{1}{\sqrt{8}\pi l_s} \,.
\end{equation}
The reason that one expects this to be the Hagedorn temperature for large 't Hooft coupling is as follows. 
Consider a string with energy $l_s E$ in string units. If this energy is sufficiently high, also compared to the angular momenta on AdS$_5 \times S^5$, a particle mode of the string is probing distances shorter than the radius $R$ of AdS$_5$ and $S^5$. Moreover, if $\sqrt{l_s E} \ll R/l_s$, the extension of the string is much smaller than the radius $R$. Thus,  there is an intermediate regime in which the spectrum of a string behaves as in flat space. As $\lambda\rightarrow\infty$, this means that the Hagedorn temperature approaches that of flat space.

\section{Conclusion and Outlook}

In this Letter, we have derived integrability-based QSC equations that determine the Hagedorn temperature of planar $\mathcal{N}=4$ SYM theory -- and equally of type IIB string theory on AdS$_5\times S^5$ -- at any value of the 't Hooft coupling.
We have solved these equations perturbatively at weak coupling, reproducing known results up to two-loop order and obtaining the previously unknown three-loop result \eqref{eq: perturbative result}.
The same algorithm can also be used at higher orders, as we will demonstrate in an upcoming publication \cite{Harmark:2021qma}.
Moreover, we have solved the QSC numerically at finite coupling, allowing for an interpolation between weak and strong coupling.
From our numeric results, we have read off the first coefficient in the strong coupling expansion.
It would be interesting to increase the numeric precision even further using a C$^{++}$ implementation following Ref.\ \cite{Hegedus:2016eop} to obtain further coefficients in the strong coupling expansion.

We have found evidence that the Hagedorn temperature, which marks the temperature beyond which the planar partition function is singular, asymptotes to the Hagedorn temperature of type IIB string theory in ten-dimensional Minkowski space for large $g$. This is in line with the expectation that for large $g$ the spectrum should approach that of type IIB string theory on flat space, as explained above. 
To test further that the spectrum approaches flat space, one could possibly use the techniques of this Letter to study the critical behavior of the partition function close to the Hagedorn singularity for large $g$.
If the critical behavior matches the one of flat-space string theory, it would confirm that there is a regime of strongly coupled $\CN=4$ SYM theory in which the spectrum is that of tree-level type IIB string theory in ten-dimensional Minkowski space. 
Thus, this Letter opens up an interesting new regime in which one can explore the AdS$_5$/CFT$_4$ correspondence.

Note finally that while we have restricted ourselves to vanishing chemical potentials, the case of non-vanishing chemical potentials can be treated in a similar way \cite{Harmark:2021qma}. 
This could provide a connection to the cases of the Hagedorn temperature in the pp-wave or spin-matrix-theory limits \cite{PandoZayas:2002hh,Greene:2002cd,Brower:2002zx,Grignani:2003cs,Harmark:2006ta,Harmark:2014mpa,Yamada:2006rx,Harmark:2006di,Suzuki:2017ipd,GomezReino:2005bq}.
Moreover, it would be interesting to consider the Hagedorn temperature for integrable deformations of $\CN=4$ SYM theory (see \cite{Fokken:2014moa} for one-loop results) and for the three-dimensional $\CN=6$ superconformal Chern-Simons theory, for which a QSC formulation for the spectral problem exists as well \cite{Bombardelli:2017vhk}. In particular, it would be intriguing to study what happens at strong coupling in these cases.

\begin{acknowledgments}
\paragraph{Acknowledgements}
%
We thank
Simon Caron-Huot,
Marius de Leeuw,
Claude Duhr,
Nikolay Gromov,
Sebastien Leurent,
Fedor Levkovich-Maslyuk,
Christian Marboe,
Andrew McLeod,
Stijn van Tongeren,
Matt von Hippel
and Konstantin Zarembo
for very useful discussions. 
We thank Nikolay Gromov for sharing his Mathematica implementation of the algorithm in Ref.\ \cite{Gromov:2015wca}.
M.W.\ thanks the Institute for Advanced Study in Princeton for kind hospitality.
T.H.\ acknowledges support from FNU grant number DFF-6108-00340.
M.W.\ was supported in part by FNU through grant number DFF-4002-00037, the ERC starting grant number 757978, the Danish National Research Foundation (grant number DNRF91) and the Villum Fonden.
\end{acknowledgments}

\bibliography{mybib}

\end{document}